\documentclass{aa}
\usepackage{graphicx}
\usepackage{amssymb}
\usepackage{amsmath}
\newcommand{\etal}{et al.}

\newcommand{\muKrs}{~\mu{\rm K}\sqrt{{\rm s}}}
\newcommand{\apj}{ApJ}
\newcommand{\PR}{Phys. Rev.}
\newcommand{\AsAs}{A\&A}

\newcommand{\MNRAS}{MNRAS}

\newcommand{\Anst}{Ann. Statist.}
\newcommand{\rb}[1]{\raisebox{1.5ex}[0pt]{#1}}
\begin{document}
\title{Non-iterative methods to estimate the in-flight noise properties
of CMB detectors}
 
\subtitle{}
\author{P. Natoli\inst{1} \and D. Marinucci\inst{2} \and P. Cabella\inst{1} 
\and G. de Gasperis\inst{1} \and N. Vittorio\inst{1}}

\institute{Dipartimento di Fisica, Universit\`a di Roma ``Tor Vergata'', 
via della Ricerca Scientifica 1, I-00133, Roma, Italy \and 
Dipartimento di Studi Geo\-economici e Sta\-tistici,
Universit\`a di Roma ``La Sa\-pien\-za'', via del Castro Laurenziano 9, I-00161
Roma}

\offprints{Paolo.Natoli@roma2.infn.it}

\date{Received / Accepted}

\abstract{
We present a new approach for statistical inference on
noise properties of CMB anisotropy data. We consider
a Maximum Likelihood parametric estimator to recover the
full dependence structure of the noise process. We also
consider a semiparametric procedure which is only sensitive
to the low frequency behavior of the noise spectral density.
Both approaches are statistically robust and computationally
convenient in the case of long memory noise, even under
nonstationary circumstances. We show that noise
properties can be consistently derived by such procedures without resorting
to currently used iterative noise-signal methods. More importantly,
we show that optimal (GLS) CMB maps can be obtained from
the observed timestream with the only knowledge of the noise
memory parameter, the outcome of our estimators.
}

\titlerunning{Non-iterative methods to estimate the in-flight\ldots}

\maketitle
\section{Introduction}\label{intro}

Recent high sensitivity, high resolution observations of the Cosmic
Micro\-wave Back\-ground (CMB) anisotropy are posing
strong constraints on Cosmology (see, e.g., 
Balbi et al., 2000; de Bernardis et al., 2001).
The ability to discriminate among 
different cosmological scenarios critically
relies on the efficiency and accuracy of the data analysis pipeline, other
than -obviously- on experimental know-how. 
In order to be efficient, data analysis has to deal with the size of the 
data sets, which is bound to increase with new space missions such as 
MAP\footnote{http://map.gsfc.nasa.gov/} and 
{\sc Planck} Surveyor\footnote{
http://astro.estec.esa.nl/SA-general/Projects/Planck/}.
In order to be accurate, the data analysis pipeline should implement
optimal statistical techniques to reconstruct the correlation structure of
receiver noise, i.e.\ the behavior of its spectral density. 
This is a critical point for 
data analysis of CMB anisotropy one-horned experiment. In fact, both
Maximum Likelihood map making 
(see e.g. Wright, 1996; Borrill, 1999; Natoli et al., 2001) 
and angular power spectrum estimation (Tegmark 1997; Bond et al., 
1998; Oh et al., 1999) 
heavily depend on the knowledge of the noise spectral density, which
of course must be determined from the data themselves.

Our paper is not the first to focus on estimating the 
noise properties out of a CMB anisotropy data stream. 
Ferreira \& Jaffe (2000) advocated 
simultaneous estimation of signal and noise on the basis of a 
Bayesian approach. However, their estimates for the noise spectral density
are piecewise constant with several discontinuities. Far from being
realistic, this assumption does not apply to long
range dependence models.
Prunet et al.\ (2001) obtained a purely nonparametric 
estimate of the noise spectral density as a by-product of 
their iterative map making procedure. Even if cast in a frequentist
framework, the resulting algorithm is very similar to
Ferreira and Jaffe's proposal. It is worth noting that, in both cases, 
purely nonparametric estimates of noise properties
lack a rigorous justification in the statistical literature if 
noise is a long memory process. In this case,
all of these difficulties
can be overcome by using well known and robust statistical techniques
which are well supported by the most recent literature
(Fox \& Taqqu, 1986; Dahlhaus, 1995;
Robinson 1995ab; Giraitis \& Taqqu, 1999).
In Sect.~2 we discuss these techniques in the context
of CMB data analysis. In Sect.~3 we apply them to simulated datasets
of the {\sc Planck} and 
BOOOMERanG\footnote{http://www.physics.ucsb.edu/~boomerang/}
experiments. Finally, in Sect.~4 we draw
our main conclusions and point out directions for future research.

\section{Efficient estimation of the noise properties}

We assume that the data, ${\bf d}$, consist of the sum of two 
uncorrelated time series:
\begin{equation}\label{unc_sum}
d_t = s_t + n_t
\end{equation}
for $t=1,2,\dots , T$. Here ${\bf n}$ and  ${\bf s}$ are noise
and signal, respectively.
The latter is modeled assuming linear dependence on the sky pattern:
\begin{equation}\label{linear_model}
s_t = \sum_{p=1}^{K} P_{tp} m_{p}. 
\end{equation}
We imagine the ``map''  ${\bf m}$ to be smeared by the instrumental beam
and discretized into $K$ elements (or pixels). The matrix ${\bf P}$
effectively defines the experimental scanning strategy
by ``unrolling'' the sky map into a time series. We limit ourselves
to the case of one-horned experiments and symmetric beam profile, 
for which the matrix ${\bf P}$
has a very simple structure, consisting of a single non zero entry (with
value $1$) per row. This entry corresponds to the pixel 
observed at a given time $t$.
For very low signal to noise (S/N) ratio the timestream, ${\bf d}$,
is noise dominated. In this Section we pretend that this is the case.
We will show below, in Sect.~3, how to drop this assumption.

\subsection{Long memory processes}
A process is said to be wide sense (or second order) stationary if
its mean, variance and covariance functions are finite and constant over
time: $\langle n_{t}\rangle=\mu$, $\langle n_{t}n_{t}\rangle=
\gamma (0)<\infty$ and 
$\langle n_{t}n_{t-\tau }\rangle=\gamma (\tau )$, $\tau =\pm 1,\pm 2,\dots$.
For a noise process we shall take $\mu \equiv 0$.
The Cram\'{e}r Representation Theorem ensures that for any second order
stationary process we can write:
\begin{equation}
n_{t}=\int_{-\pi }^{\pi }\exp (itf)\sqrt{P(f)}\,dW(f)\, ,
\label{srt}
\end{equation}
where $W(f)$ is a random measure such that $\langle dW(f) \rangle =0$, while 
$\langle dW(f)dW(g)\rangle = df$ if $f=g$ and $0$ otherwise (i.e. 
$W$ represents a white noise measure). 
The function $P(f)$ 
is the spectral density function of the process ${\bf n}$,
uniquely related to the sequence of covariances $\gamma (\tau)$ through
the usual Fourier formulae:
\begin{eqnarray*}
P(f) &=&\frac{1}{2\pi }\sum_{\tau =-\infty }^{\infty }\gamma (\tau )\exp
(-i\tau f)\text{ , } \\
\gamma (\tau ) &=&\int_{-\pi }^{\pi }\exp (-i\tau f)P(f)df\text{ .}
\end{eqnarray*}
Thus, wide sense stationarity implies: 
\[
\gamma (0) = \int_{-\pi }^{\pi }P(f)df<\infty \text{ ,}
\]
that is, 
the spectral density needs always be integrable.
State of the art microwave detectors produce ``$1/f$'' noise.
We shall therefore focus on long memory (or long range
dependent) noise. For this processes the
spectral density is such that
\begin{equation}\label{lmdef}
\lim_{f \rightarrow 0}\frac{P(f)}{f^{-\alpha }}=G,
\end{equation}
where $0<G<\infty$ and $0<\alpha <1$.
So, even if divergent at zero frequency, the spectral density
is always integrable. Eq.~(\ref{srt}) can be discretized as follows:
\begin{equation}
n_{t}=\frac{1}{\sqrt{2\pi T}}\sum_{j=1}^{T}\exp (itf_{j})\sqrt{P(f_{j})}\;
\varepsilon _{j},  \label{srtsim}
\end{equation}%
Here $f_{j}=2\pi j/T$ are Fourier frequencies, while $\varepsilon _{j}$
denotes a sequence of zero-mean, independent and
identically distributed ($i.i.d.$) variables with unit variance.
Eq.~(\ref{srtsim}) is widely used to generate random processes with a given 
spectral density $P(f)$ in
Monte Carlo procedures. If we take, for instance, $P(f)=Gf^{-\alpha }$,
straightforward algebra shows that:
\[
\lim_{T\rightarrow \infty} \langle n_t n_t \rangle
\simeq \lim_{T\rightarrow \infty}\frac{1}{2\pi T} \sum_{j=1}^{T}P(f_j) =
\left\{ 
\begin{array}{c}
O(1),\text{ }\alpha <1 \\ 
O(\log T),\text{ }\alpha =1 \\ 
O(T^{\alpha -1}),\text{ }\alpha >1%
\end{array}
\right. .
\]
Therefore, if $\alpha \ge 1$ the variance of ${\bf n}$ grows with time,
i.e.\ the process is not stationary.
We shall 
first focus on the stationary case $\alpha <1$. This
assumption will be relaxed to $\alpha <2$ in Sect.~\ref{non_stat_case}.

\subsection{Fully parametric procedures}

Assume first that the the functional form of the 
spectral density for the process ${\bf n}$ is known and
completely specified by a set of parameters 
${\bf \theta} \equiv (\theta_1, \theta_2, \dots, \theta_N)$.
For ``$1/f$'' noise it is customary to write:
\begin{equation}\label{planck_noise_model}
P(f,{\bf \theta})=A\left\{ 1+\left (\frac{f_k}{f} \right )^{\alpha }\right\}
\end{equation}
$\theta =(A,\alpha,f_{k})$, where $A$ is an amplitude, $f_k$ is 
the so-called knee frequency and $\alpha$ is the 
spectral index introduced above.
Throughout this paper we shall assume that  ${\bf n}$ is Gaussian
even if this condition is not essential for our conclusions to
hold. Under Gaussianity ${\bf n}$ has a multivariate density
function: 
\begin{equation}
\Psi(\mathbf{n})=
\frac{1}{\sqrt{(2\pi )^{T} |\mathbf {N}| }}\exp \left\{ -\frac{1}{2}%
\mathbf{n}^{\prime }\mathbf {N} ^{-1}\mathbf{n}\right\}\, ,  \label{df}
\end{equation}
where $\mathbf {N}$ is the noise covariance matrix 
and the prime ($^{\prime }$) denotes transposition.
Now, for large $T$ the matrix $\mathbf{N}^{-1}$ is nearly circulant and
can be approximated as: 
\begin{equation}\label{circapp}
\mathbf {N}^{-1} \simeq Q^{\dagger}\Lambda^{-1}Q 
\end{equation}
where $\Lambda =diag\{P(f_{1}),...P(f_{T/2})\}$,
the dagger ($\dagger$) denotes complex conjugation 
coupled with transposition and $Q$ is the rectangular 
($T/2 \times T$) matrix
\begin{equation}
Q_{jt} = \frac{e^{itf_j}}{\sqrt{2\pi T}}\, .
\end{equation}
The sample Discrete Fourier Transform (DFT)
\[ 
Q\,\mathbf{n}\equiv\{w_{T}(f_{1}),\dots,w_{T}(f_{T/2})\}\, ,
\]
has elements
\[
w_{T}(f_{j})=\frac{1}{\sqrt{2\pi T}}
\sum_{t=1}^{T}n_{t}e^{itf_{j}}\,.
\]
The sample periodogram is defined as 
\[ 
I_{T}(f_{j})=w_{T}(f_{j})w_{T} (f_{j})^{\ast}\;, 
\]
where the asterisk denotes conjugation.
 
From Eq.~(\ref{df}), it is immediate to derive the 
log-likelihood function for $\theta $ as 
\[
\mathcal{L}(\theta ;\mathbf{n})=-\frac{T}{2}\log 2\pi -\frac{1}{2}\log 
|\mathbf{N} (\theta )|-%
\frac{1}{2}\mathbf{n}^{\prime }\mathbf{N}(\theta )^{-1}\mathbf{n}\;.
\]
Maximization of this form in the time domain is a computationally unfeasible
task even for moderately large values of $T$, let aside 
$T \gg 10^{7}$ that will become customary in CMB experiments. 
Based only on the approximations in Eq.~(\ref{circapp})
and building upon ideas that trace back to
work by Whittle (1953), Fox and Taqqu (1986) 
have suggested the following frequency domain 
expression, valid up to a constant term:
\begin{eqnarray}
-2\mathcal{L}(\theta ;\mathbf{n}) &\simeq& \log |Q\Lambda Q^{\dagger
}|+\mathbf{n}^{\prime }Q^{\dagger }\Lambda^{-1}Q\mathbf{n} \notag \\
&=&\log |Q^{\dagger }Q|+\log |\Lambda| 
+(Q\mathbf{n})^{\dagger }\Lambda^{-1}Q\mathbf{n} \notag  \\
&=&\sum_{j=1}^{T/2}\log P(f_{j};\theta ) 
+ \sum_{j=1}^{T/2}\frac{I_{T}(f_{j})}{P(f_{j};\theta )}.
\label{whi}
\end{eqnarray}
The Whittle Estimate (WE) for the noise parameters, $\theta$, 
is then defined as:
\[
\widehat{\theta}=\arg \min_{\theta} \left\{
-\sum_{j=1}^{T/2}\log P(f_{j};\theta )-\sum_{j=1}^{T/2}\frac{I_{T}(f_{j})}
{P(f_{j};\theta )}\right\},
\]
where $\arg \min_{x} \{f(x)\}$ denotes the 
value $x$ which minimizes $f(\cdot)$. 
The same authors show that, under regularity conditions, these 
estimates are $\sqrt{T}$--consistent, asymptotically unbiased and Gaussian,
meaning that
\begin{equation}\label{asymp_distr}
\lim_{T\rightarrow \infty}\sqrt{T}\;(\widehat{\theta } - \theta) \sim
\mathcal{N}(0,\Sigma)
\end{equation}
where $\mathcal{N}$ denotes a Gaussian multivariate with covariance matrix
$\Sigma$ whose explicit expression is given by Fox and Taqqu (1986).
The estimate
$\widehat{\theta }$ is computationally very convenient and Dahlhaus
(1989) proves that it is asymptotically equivalent to the corresponding
estimate in the time domain, i.e.\ to 
standard Maximum Likelihood. It follows
immediately that the 
WE is absolutely efficient in the Cram\'{e}r-Rao 
sense, that is it  achieves minimum variance. 
Generalizations to a non Gaussian framework are considered by
Giraitis and Surgailis (1990) and Giraitis and Taqqu (1999).
Thus,
provided that the functional 
form of the spectral density is known {\it a priori},
WE solves the problem of optimal inference of the noise
properties given that the observed timestream is noise dominated.

\subsection{Semiparametric procedures}
In some situations the functional form of the spectral density
may not be known. Nonetheless, in the presence of long range dependence, 
knowledge of the low frequency behavior of the noise 
spectral density is all that is needed to implement 
optimal filtering procedures (Dahlhaus, 1995). Thus, determining
$\alpha$  is sufficient for Generalized Least Squares (GLS) or Maximum
Likelihood (ML) map making.
This fact calls for statistical methods which, rather than parameterize
the full spectral density, only rely on the much milder condition of
Eq.~(\ref{lmdef}).
One such procedure, the 
so-called Log Periodogram Estimate (LPE), was introduced by Geweke and
Porter-Hudak (1983) and discussed rigorously by Robinson (1995a). 
Consider the identity: 
\begin{equation}\label{regr_model}
\log I_{T}(f_{j})=\log G-\alpha \log f_{j}+\log \frac{P(f_{j})}{%
Gf_{j}^{-\alpha }}+\log \frac{I_{T}(f_{j})}{P(f_{j})}\, ,
\end{equation}
where $j=1,2,\dots,m<T/2$ and $m = m(T)$ is chosen such that 
$\lim_{T\rightarrow\infty}1/m = 0$ and
$\lim_{T\rightarrow\infty}m/T = 0$.
Under this Asymptotic Bandwidth Condition (ABC) and in 
view of Eq.~(\ref{lmdef}),
one has that $\log [P(f_{j})/Gf_{j}^{-\alpha }]\rightarrow 0$ 
whereas $\log [I_{T}(f_{j})/P(f_{j})]$ can be approximated
as a sequence of zero-mean and nearly uncorrelated residuals, with finite
variance. Therefore, it seems natural to consider Eq.~({\ref{regr_model})
as a regression model. 
This heuristic is made rigorous by Robinson (1995a), where the
Ordinary Least Square (OLS) estimate is considered:
\begin{equation}
\widehat{\alpha }=\arg \min_{\alpha ,G}\sum_{j=1}^{m}\left\{ \log
I_{T}(f_{j})-G-\alpha \log f_{j}\right\} ^{2}\text{ .}  \label{logper}
\end{equation}%
The intuitive meaning of the ABC 
is that we only consider a vanishing
(as $T \rightarrow \infty$) subset of frequencies around the origin.
Because of this, 
we are actually discarding most of the available information as it is
the case for semiparametric procedures. The practical consequence,
highlighted in Robinson (1995a), is that $\widehat{\alpha}$, despite
being asymptotically Gaussian and unbiased, is 
only $\sqrt{m}$-consistent (in
the sense of Eq.~(\ref{asymp_distr})).
It has therefore asymptotical efficiency zero with respect to
WE when the spectral density is correctly parameterized. In principle,
a misspecified model may entail inconsistent
estimates of all parameters, and in particular of $\alpha$, which is 
the parameter of interest for many, if
not most, applications. In this context, robust semiparametric procedures
may represent a more reliable choice.

Robinson (1995b) considers another semiparametric
method which can be viewed as a narrow-band analogous of
WE\@. Its properties and
motivating rationale, however, are too close 
to LPE to warrant independent consideration in this paper.
 
A somewhat intermediate attitude between the fully parametric WE
and the semiparametric LPE has been recently set forth
by Moulines \& Soulier (1999). The idea is
basically to use Eq.~(\ref{lmdef}) to analyze the lowest
frequencies, where the long range dependent behavior shows up, and to
consider a series expansion into orthogonal components for the remaining
part of the spectral density. 
These estimates are in principle appealing because they can be shown to
converge as fast as $\sqrt{T/\log T}$, provided that the spectral density 
$P(f)$ is sufficiently well behaved close to the origin 
(see Moulines \& Soulier, 1999) for more details). 
However, they are computationally more costly than WE and LPE
because they require multiple regression with hundreds or
thousands of estimated explanatory terms. 

\subsection{The nonstationary case}\label{non_stat_case}

The analysis in the previous Section assumes
stationarity of the noise series. This assumption
may turn out to be too strong over a long time span.
For instance, the variance of the noise
might grow (slowly) as the observation goes on. A general
model for such a process might be 
\[
n_{t}=\sum_{j=0}^{t}\psi _{j}\:\varepsilon _{t-j},
\]
where $\varepsilon _{t}$ represents a sequence of 
independent innovations with zero mean and finite variance, 
whereas the weight sequence $\psi _{j} \simeq \kappa j^{\alpha /2-1}$, 
($\kappa > 0$). 
For $\alpha < 1$, it is not
difficult to see that $\mathbf{n}$ is asymptotically stationary 
and satisfies Eq.~(\ref{lmdef}) 
as $t\rightarrow \infty$. For $1<\alpha <2$ Velasco (1999)
proves that, although the spectral
density is no longer defined, the expected value of the periodogram maintains
the same behavior as in Eq.~(\ref{lmdef}). In the case $\alpha =2$,  
$\mathbf{n}$ behaves (asymptotically) as a pure random 
walk process:
\[
n_{t}\simeq \kappa\sum_{j=0}^{t}\varepsilon _{j}\text{ .}
\]
The case $\alpha >2$ can also be considered. Robinson \& Marinucci
(2001) investigate the behavior of the periodogram for any positive 
value of $\alpha$. However, this would entail a 
superlinear growth of the variance with time. We rule this
out as experimentally questionable. 

LPE and WE are shown to yield consistent 
estimates even under the condition $\alpha < 2$
(Velasco, 1999; Robinson \& Velasco, 2000). This fact allows us
to also consider values of $\alpha$ greater than unity in
the simulations we will carry out in the next Section. 
As a consequence,
the procedures
advocated in this paper enjoy a marked advantage with respect to those so
far considered in the CMB literature, 
to the extent in which the latter analyses
are deeply rooted in a stationary framework and apparently lack
rigorous justification otherwise.

\section{Numerical Applications}
The methods outlined above assume that the noise time series,
$\mathbf{n}$, is known. Obviously this is not the case for a real world
experiment where the observed sample is a combination of noise
and signal. This is precisely the reason why iterative methods
have grown attention
in the literature. In this paper, however, we will follow 
a different -- yet simpler -- approach.
That is, we will test the efficiency
of the statistical methods outlined above on the simplest noise
estimator, obtained by subtracting the OLS (i.e.\ naively
coadded) map from the data. To be more specific, let us write 
Eq.~(\ref{unc_sum}) in matrix form as:
\[
{\bf d} = {\bf P}{\bf m} + {\bf n}.
\]
The OLS estimator is then:
\begin{equation}\label{n_tilde} 
{\tilde {\bf n} } 
\equiv \mathbf{d} - \mathbf{P} (\mathbf{P}^T \mathbf{P})^{-1}
\mathbf{P}^T \mathbf{d} = [\mathbf{I}-\mathbf{P} 
(\mathbf{P}^T\mathbf{P})^{-1}\mathbf{ P}^T]\mathbf{n}
\end{equation}
which does not include any contribution from the 
signal, independently of the specific S/N ratio
of the timestream (Natoli et al.\ 2001).

The plan of this Section is as follows.
First, we test the efficiency of WE and
LPE on the pure noise ($\mathbf{n}$)
timestream. We then compare these results with those obtained by
using the ${\tilde {\bf n}}$ 
estimator. To assess this point, we make Monte Carlo
simulations for the {\sc Planck} mission. 
We then use our noise estimates to perform
map making on simulated datasets for both {\sc Planck} and BOOMERanG. 
In doing so, we also compare the outcome of our noise
estimation method with our implementation of the
iterative scheme proposed by Prunet et al.\ (2001)
for BOOMERanG.

\subsection{LPE and WE efficiency on $\bf n$ 
and ${\tilde {\bf n} }$}\label{app_to_planck}

Our first test bed is 
the 30~GHz channel of {\sc Planck}/LFI. We remind here that {\sc Planck} spins
at 1~rpm, has a boresight angle of  $85^\circ$ and
observes the same circle of the sky for 1 hour 
(every hour the spin axis is moved, say along the ecliptic, by $2.5'$). 
The mission is planned to remain in operation for at least 14 months, resulting
in approximately two nearly full sky coverages. This corresponds to
$\sim 10^9$ samples for each of the two LFI 30~GHz horns. 
The size of this
timestream makes Monte Carlo simulations quite prohibitive. 
Therefore, we have decided to  use only $\sim 2^{24}$ ($\sim 2\cdot 10^{7}$) 
observations corresponding to $\sim 140$ rings (each ring is scanned 60
times) and covering two slices, each 
about $12^\circ$ degrees wide. This means that we process
$\sim 12$ days of observation. 
For the moment, we choose the usual ``$1/f$'' 
noise given in Eq.~(\ref{planck_noise_model}).
As discussed in Sect.~2, we will examine both the $\alpha < 1$
and the $\alpha \geq 1$ (i.e., non stationary) case. The values
of $A$ and $f_k$ are chosen accordingly to the {\sc Planck} specifications:
$A \simeq 4\cdot 10^4\,(\muKrs)^2$ and $f_k = 0.1$~Hz.
The simulations are performed by first generating a pure noise
timestream, ${\bf n}$, according to Eq.~(\ref{srtsim})
with the power spectrum given in Eq.~(\ref{planck_noise_model}).
This is accomplished by use of standard FFT techniques.
The ${\tilde {\bf n}}$ estimator is trivially computed using
Eq.~(\ref{n_tilde}).
The periodograms of the series
are then estimated by binning together a given number of 
neighboring frequencies
(in the statistical literature this procedure is known as ``pooling
of the periodogram''). This choice has two
advantages: (1) it brings down the size of the dataset (from $2^{12}$
proper frequencies to, say, a few thousands) while preserving most,
if not all, of the relevant information; (2) it is potentially
beneficial from the point
of view of the efficiency of the estimator (see Robinson, 1995a
for discussion concerning the LPE case). 
Simulations show that WE and LPE
are affected, though not dramatically, by the number of frequencies
pooled together; in fact, if we bin too many frequencies
we can even degrade the quality of the estimator. 
We found that a reasonable choice is to pool together 
$64$ frequencies.
As discussed in Sect.~2, LPE only exploits
the $m$ lowest frequencies of the periodogram. 
In our simulations, we find that
an optimal number is $m \simeq 80$.\footnote{ 
In the statistical literature it is occasionally suggested to trim 
(i.e.\ drop) the lowest ($\sim 10$) frequencies of the periodogram. We
have verified that, in our case, this choice has no significant impact
on results.
}
While the implementation of LPE is straightforward, requiring
only a linear regression routine, the same
is not true for WE. In fact, in the latter case
we have to minimize Eq.~(\ref{whi}) over the chosen parameter
space which is spanned by all physically acceptable values
of $A$, $f_k$ and $\alpha$. A minimization routine is then needed. 
A seemingly good candidate is a direction set algorithm 
(see e.g.\ Powell's method described in  Press et al.\ 1992) which
is unfortunately too slow for our purposes. A much faster choice
is to use a variable metric method, such as the BFGS algorithm 
(see again Press et al. 1992). This method requires that the gradient of
the function of interest is calculated at an arbitrary point, an
information readily obtained by differentiating 
Eq.~(\ref{planck_noise_model}) w.r.t.\ the parameters.

We have performed 1000 realizations for $\alpha = 0.7$, $1$ and
$1.3$, respectively. Our results for LPE and WE are shown as 
histograms in Figs.~\ref{alfalog_alpha} to~\ref{hist_wl_wlt_1.3}
and summarized in Table~\ref{tab_WLP} where mean values and standard deviations
(across realizations) are reported. 
\begin{figure}
\resizebox{\hsize}{!}{\includegraphics{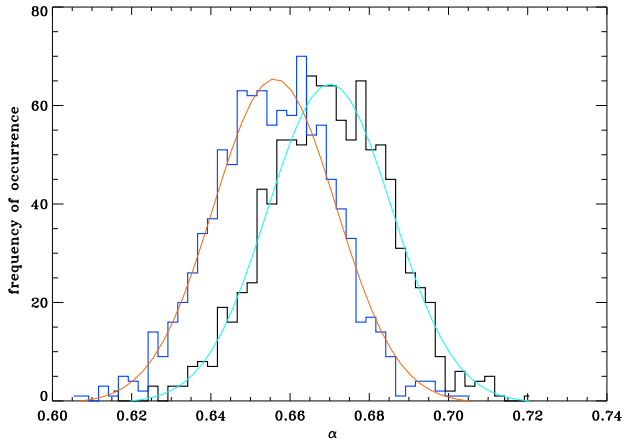}}
\caption{LP estimates for $\alpha = 0.7$ on 
${\bf n}$ (black curve) 
and ${\tilde {\bf n}}$ (dark blue curve).
The bell shaped curves overplotted are Gaussian fits to the data.}
\label{alfalog_alpha}
\end{figure}
\begin{figure}
\resizebox{\hsize}{!}{\includegraphics{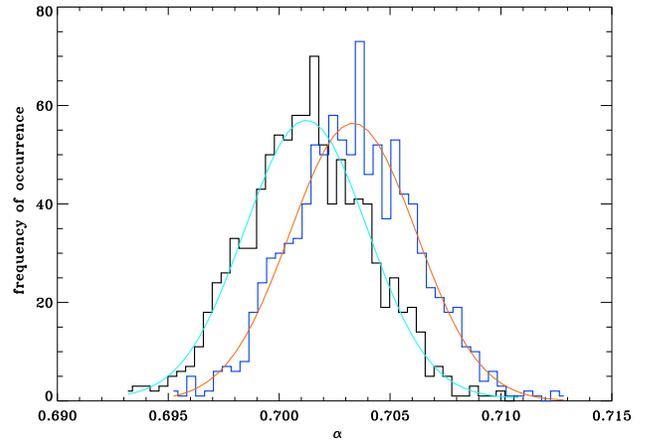}}
\caption{Same as Fig.~\ref{alfalog_alpha} but for WE.
Note the shrinking in the horizontal scale.}
\label{whittle_alfa}
\end{figure}

Our first conclusion is that both LPE and WE
are not significantly degraded when ${\tilde {\bf n}}$
is considered. In fact, the bias is at most 1\% for WE
and only slightly larger for LPE. The results for standard deviations
are also comparable. Indeed the relative performance of
LPE increases as $\alpha$ grows.
In general,
WE performs slightly better, as expected. However, the simple,
robust LPE stands up as a very convincing alternative
in the absence of precise {\it a priori} information on the functional
form of the spectral density. This point
will be further discussed below.

We have also implemented the Moulines and Soulier (1999)
broad-band log periodogram estimator
mentioned in Sect.~2. The resulting estimates, however, were
extremely close to those obtained with LPE and we thus decided
to omit their presentation for the sake of brevity. 
\begin{table*}[t]
\caption[Rec_WLP]{Recovery of relevant parameters for WE
and LPE\label{tab_WLP}}
\begin{tabular}{|c||c|c|c|c|c|c|}
\hline
 & \multicolumn{2}{|c|}{$0.7$} & \multicolumn{2}{|c|}
{$1.0$} & \multicolumn{2}{|c|}{$1.3$} \\
\rb{$\alpha_{\mathrm{input}}$} & $\mathbf{n}$ & ${\tilde {\bf n}}$ & $\mathbf{n}$ &  
${\tilde {\bf n}}$ &  $\mathbf{n}$ &  ${\tilde {\bf n}}$ \\\hline\hline
LPE & $0.671$ &  $0.657$ & $1.013$ & $0.988$ & $1.361$ & $1.314$ \\ 
\rb{${\widehat \alpha_{\mathrm{LP}}}$}
 & $\pm 0.015$ & $\pm 0.016$ & $\pm 0.017$ & $\pm 0.016$ & $\pm 0.021$ & 
$\pm 0.016$ \\\hline
WE & $0.7015$ & $0.7036$ & $1.0042$ & $1.0082$ & $1.3223$ & $1.317$ \\
\rb{${\widehat \alpha_{\mathrm{W}}}$}   
 & $\pm 0.0028$ & $\pm 0.0029$ & $\pm 0.0033$ & $\pm 0.0039$ & $\pm 0.0054$ & 
$\pm 0.030$ \\\hline\hline
\end{tabular}
\end{table*}

\begin{figure}
\resizebox{\hsize}{!}{\includegraphics{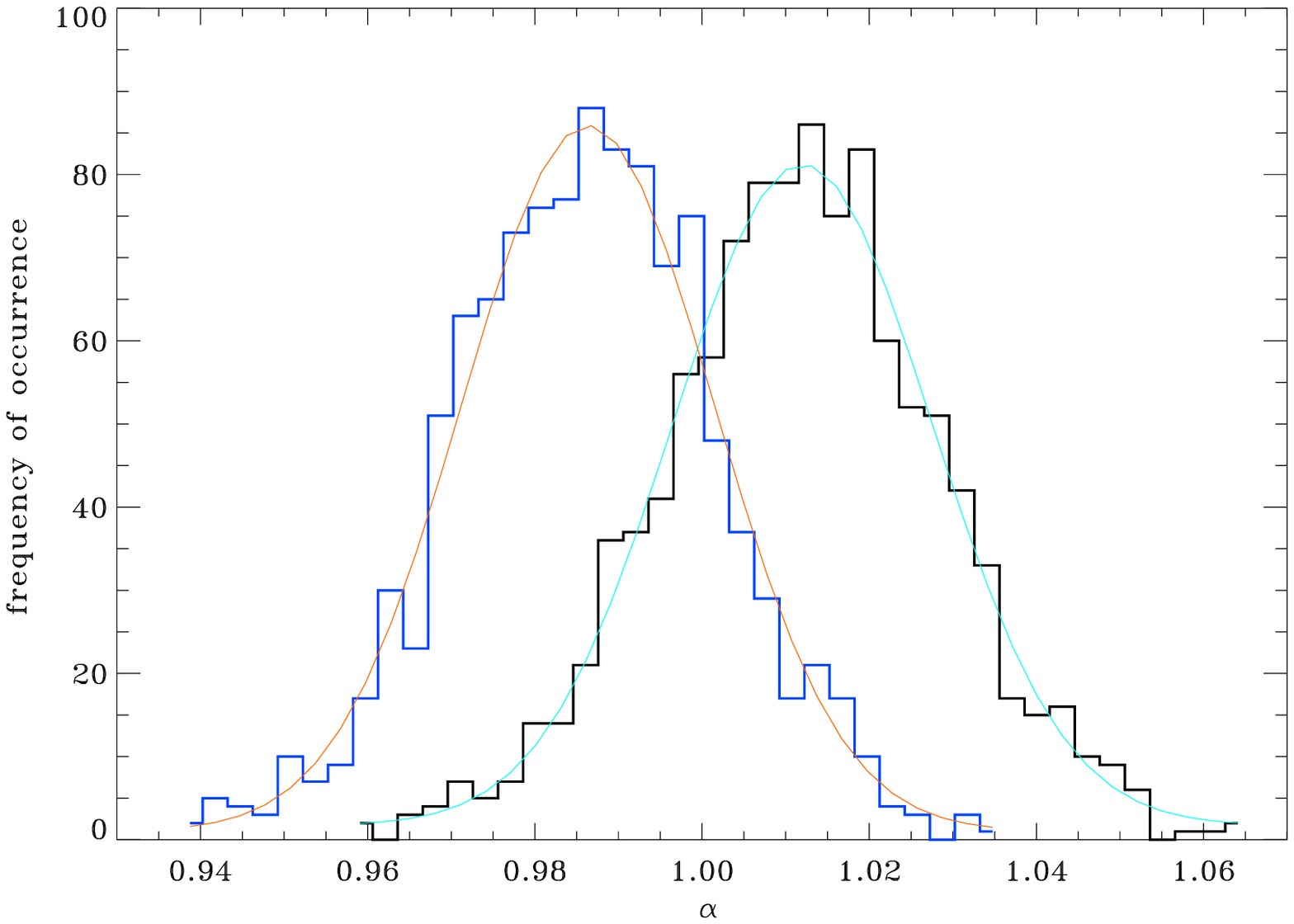}}
\caption{Same as Fig.~\ref{alfalog_alpha}, for $\alpha=1$.}
\end{figure}
\begin{figure}
\resizebox{\hsize}{!}{\includegraphics{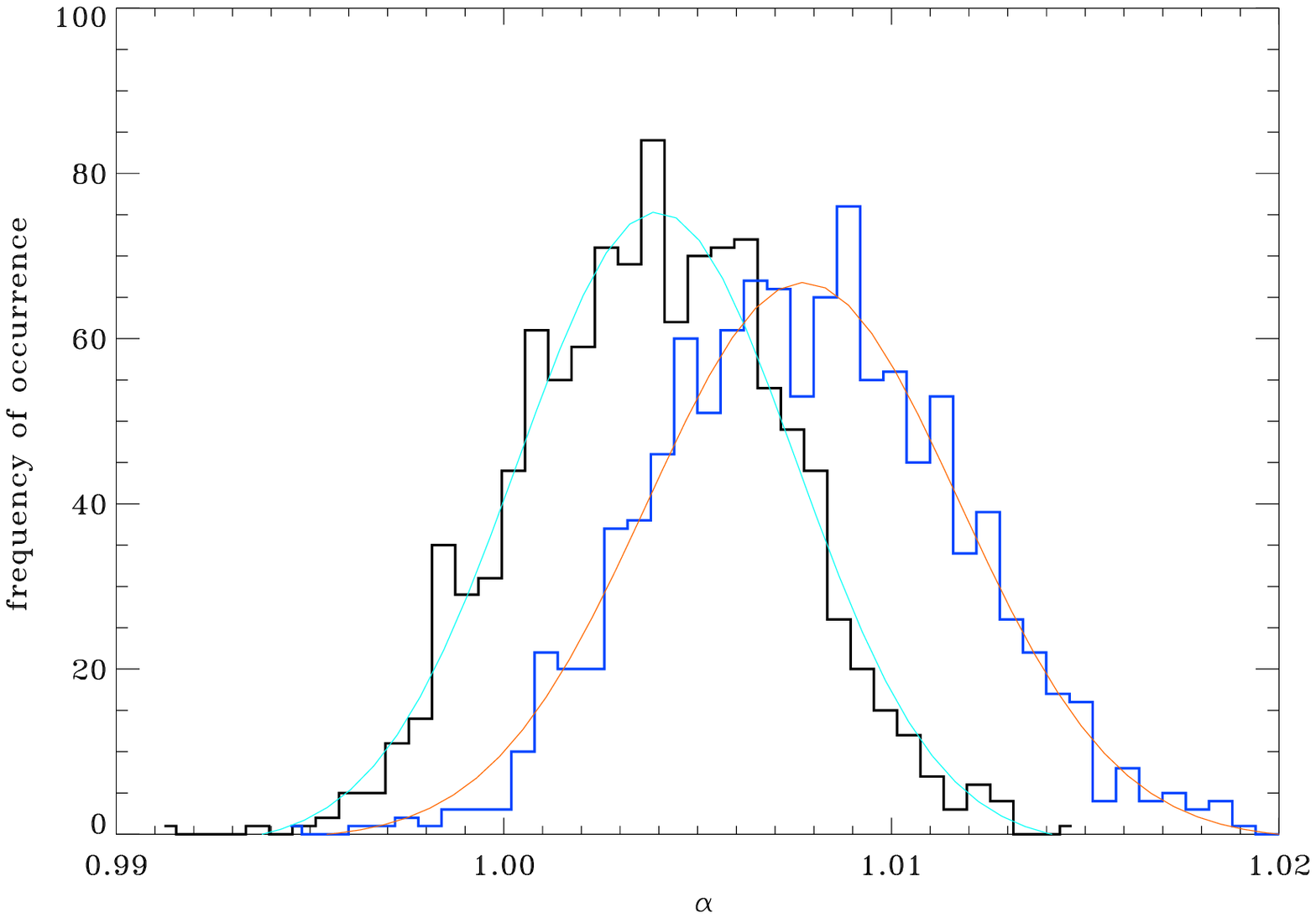}}
\caption{Same as Fig.~\ref{whittle_alfa}, for $\alpha=1$.}
\end{figure}
\begin{figure}
\resizebox{\hsize}{!}{\includegraphics{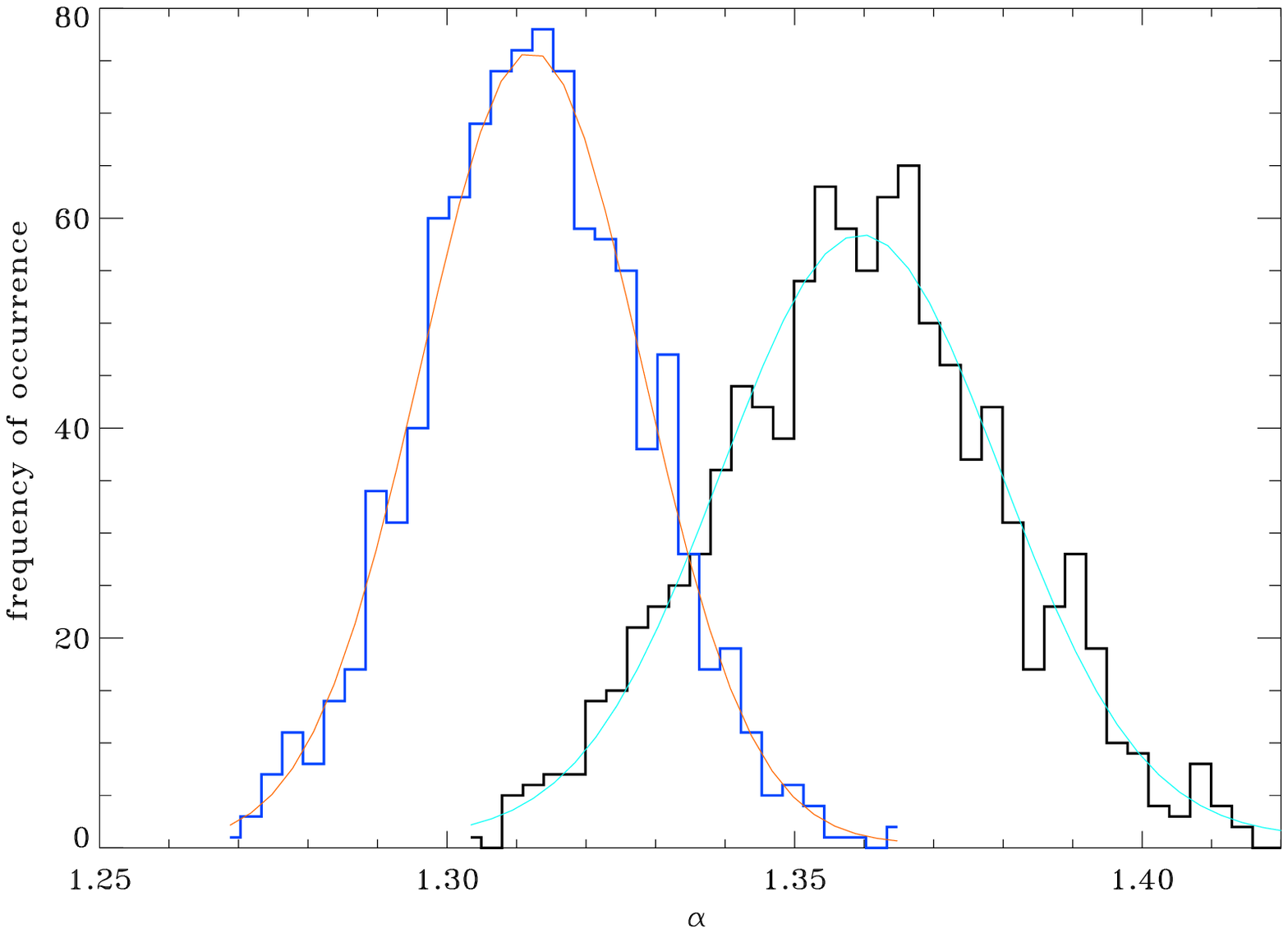}}
\caption{Same as Fig.~\ref{alfalog_alpha}, for $\alpha=1.3$.}
\end{figure}
\begin{figure}
\resizebox{\hsize}{!}{\includegraphics{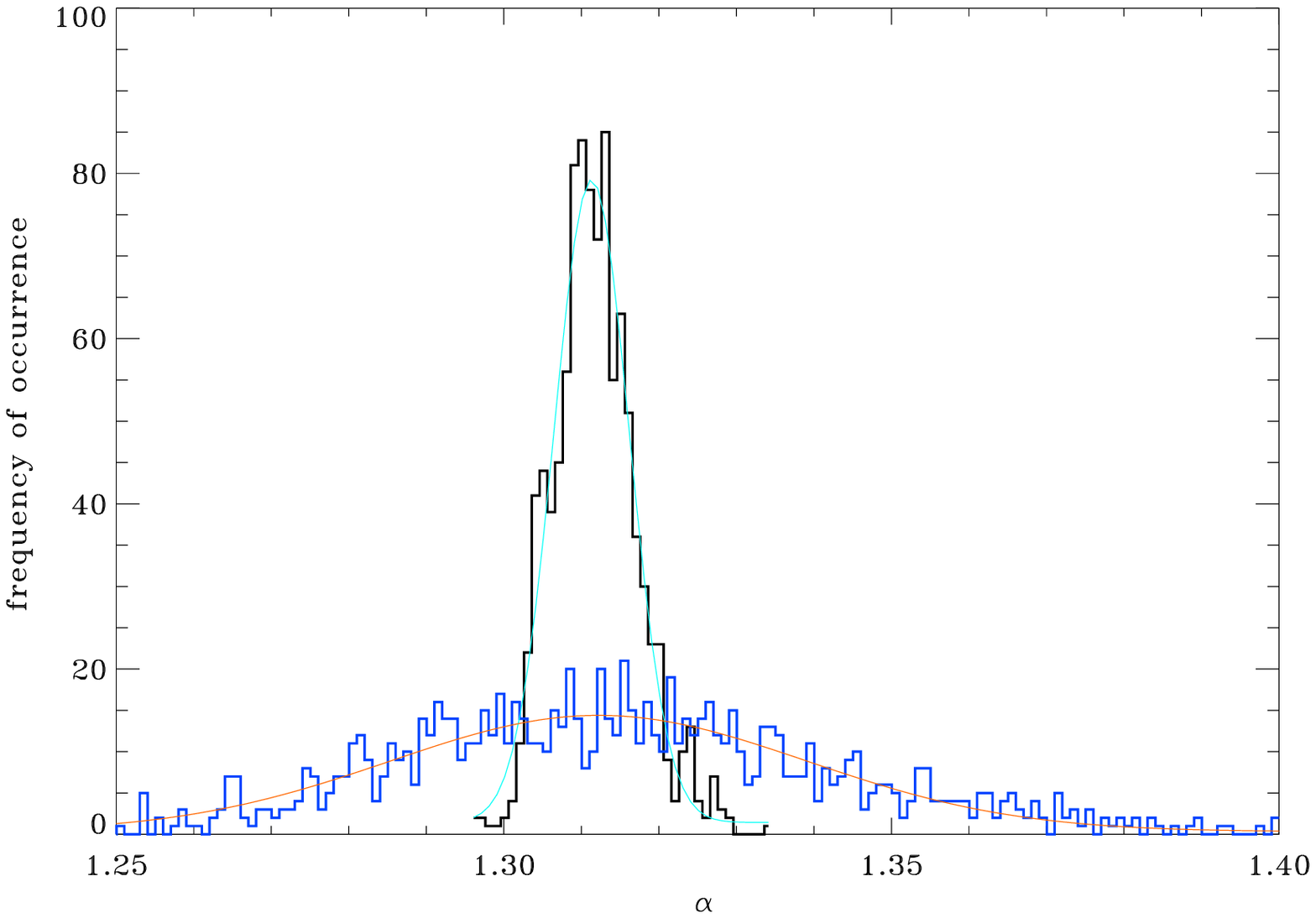}}
\caption{Same as Fig.~\ref{whittle_alfa}, for $\alpha=1.3$.
\label{hist_wl_wlt_1.3}}
\end{figure}
\subsection{Effects on Map Making}\label{lp_vs_whittle}

In order to understand how WE and LPE impact on map making
we evaluated GLS maps out of 50 data streams simulated as
discussed in Sect.~\ref{app_to_planck}. We recall here that
the GLS estimate for the map is:
\[
{\tilde {\bf m}} =({\bf P}^\prime {\bf N}^{-1} {\bf P})^{-1}  \, {\bf
P}^\prime {\bf N}^{-1} {\bf d}\, ,
\]
where $\mathbf{N}$ is the noise covariance matrix
introduced in Eq.~(\ref{df}). The Monte Carlo pipeline relies
on the following scheme: (1) generation of the noise timestream,
$\mathbf{n}$; (2) computation of ${\tilde {\bf n}}$; (3) derivation of WE
and LPE. The maps are computed using the
iterative map making algorithm discussed in Natoli et al.\ (2001). 

We note that WE return estimates of $A$, $f_k$ and $\alpha$. Only
the latter two are required to build $\mathbf{N}$, the
scale factor being irrelevant. On the other hand, LPE only returns
estimates
of $\alpha$. So, a fiducial value for $f_k$ is used
to compute ${\bf N}$. 
This approach finds its theoretical justification in
Dahlhaus (1995). He shows that taking
\begin{equation}\label{N_Dahlhaus}
N_{jk} \propto \int_{-\pi}^{\pi} e^{i(j-k)f} f^{- \alpha} 
 df  
\end{equation}
provides estimates asymptotically equivalent to exact
GLS under the only assumption that Eq.~(\ref{lmdef}) holds.
Note that Eq.~(\ref{N_Dahlhaus}) only depends on $\alpha$.

A natural figure of merit to assess the relative performance
of LPE and WE is the angular power spectrum of the maps.
The latter is computed as
\begin{equation}\label{cl_simp}
C_\ell = \frac{1}{2\ell + 1} \sum_{m=\ell}^{\ell} |a_{\ell m}|^2\, ,
\end{equation}
where the $a_{\ell m}$'s are coefficients in the spherical
harmonic expansion of the noise maps.
We are not interested here in a rigorous
power spectrum estimation but, rather, in a comparison between LPE and
WE.
This justifies the use of Eq.~(\ref{cl_simp}) which neglects
corrections for, e.g., the sky coverage and detailed shape
of the surveyed region.
The results are shown in Fig.~\ref{spec_avg_1.2} (note that
the $C_{\ell}$'s are averaged over the simulation index).
\begin{figure}
\resizebox{\hsize}{!}{\includegraphics{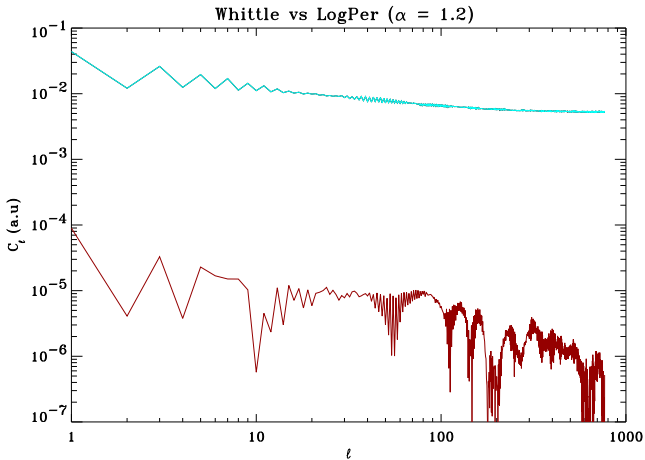}}
\caption{Averaged angular power spectrum of 50 maps displayed as
a function of multipole $\ell$. The light blue line is for WE
and completely hides the curve for LPE. The bottom red line is
the difference between the two. The spectral index input value is 
$\alpha = 1.2 $. \label{spec_avg_1.2}}
\end{figure}
The curves for WE and LPE are virtually indistinguishable.
This confirms that only the knowledge of $\alpha$ is relevant for GLS map
making. Also, the small differences between LPE and WE
become {\it de facto} immaterial when map making is at stake.
This is an important point: as discussed in Sect.~2, LPE is
semiparametric and as such
less prone to suffer from a misspecified model.

The heuristic motivation behind this result can be easily understood
if we think of GLS map making as equivalent to OLS performed on a
pre-filtered series. Pre-filtering needs not to be based on
the exact form of the noise spectral density. What really matters
is  to strongly suppress the long range correlation and this can be
achieved even with an imperfect estimate of the noise
memory parameter.

\subsection{Comparison with iterative methods}

The purpose of this Section is to compare the efficiency of LPE and WE 
against iterative noise-signal estimators. It is generally
claimed (Prunet et al.\ 2001, Dor\'e et al.\ 2001)
that, when reasonable criteria
are chosen, convergence on the noise estimator is reached fairly quickly,
and only a few ($\sim$ 4 to 5)  iterations are sufficient for most 
applications. 
 
The following scheme is employed to implement the iterative method.
We start from the OLS noise estimator, ${\tilde {\bf n} }$. 
The iteration is carried out by: (1) using the latter 
to estimate $\mathbf{N}$; (2) performing map making to obtain a
new solution for the map, ${\hat {\bf m} }$; (3) estimating
the noise again as ${\hat {\bf n} } \equiv {\bf d} - 
\bf P {\hat {\bf m} }$. This procedure, basically the same described
in Prunet et al.\ (2001) and more recently in Dor\'e et al.\ (2001),
has been implemented by using our map making tool (Natoli et al.\ 2001),
performing
6 iterations (1 OLS plus 5 GLS map making runs)
for each map, out of a total of 50 Monte Carlo realizations. 
In the same spirit of Fig.~\ref{spec_avg_1.2}
we derive $C_\ell$'s for these maps and compare them
with the results based on WE (this time for
an input value $\alpha = 0.7$). 
Results are shown in Fig.~\ref{spec_avg_0.7}:
\begin{figure}
\resizebox{\hsize}{!}{\includegraphics{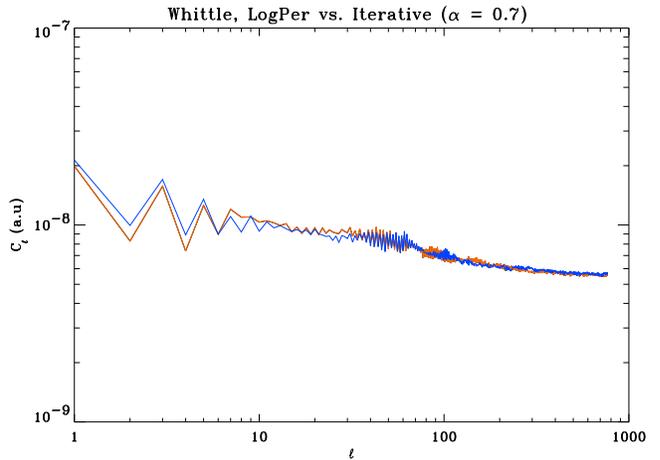}}
\caption{As for Fig.~\ref{spec_avg_1.2} but comparing
WE (blue curve) with the iterative (light red)
method. The input value for $\alpha$ is $0.7$ \label{spec_avg_0.7}}
\end{figure}
the similarity between the two curves is striking.

To further compare iterative and non iterative methods we have computed
histograms showing the frequency of occurrence of temperature
values for selected pixels of the maps. We have chosen a ``medium''
observed pixel, which is hit 660 times in the {\sc Planck} scan described
above and a ``low'' observed pixel, hit 240 times. The histograms are
shown in Figs.~\ref{hist_medobs} and~\ref{hist_lowobs}, respectively.
Note that all the pixels with the given hit number contribute
to the histograms which do not show outliers.
\begin{figure}
\resizebox{\hsize}{!}{\includegraphics{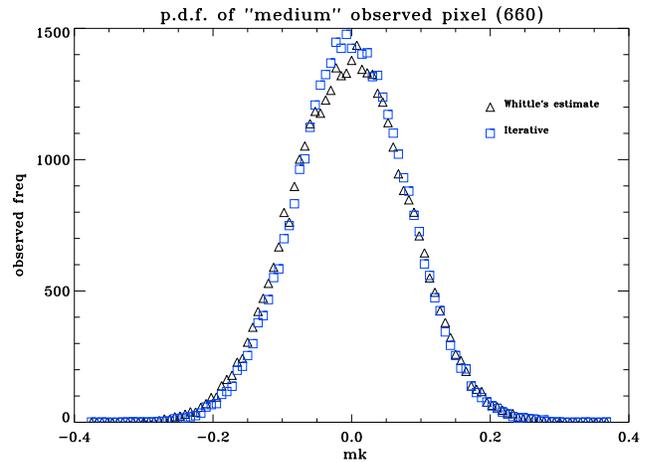}}
\caption{Empirical distribution of the ``medium'' hit pixels (660 hits).
WE (triangles) and
the iterative scheme (squares). \label{hist_medobs}}
\end{figure}
\begin{figure}
\resizebox{\hsize}{!}{\includegraphics{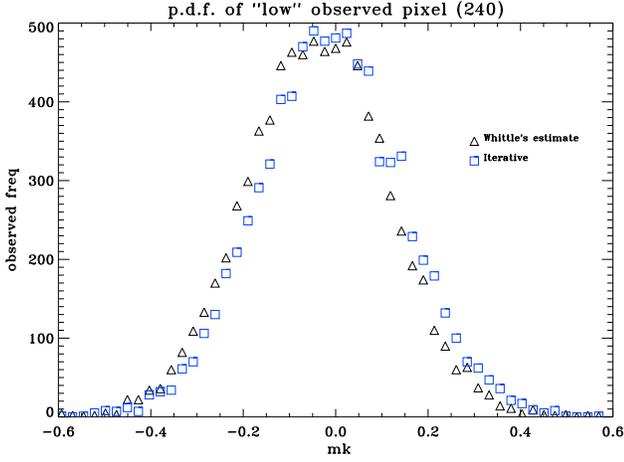}}
\caption{Same as Fig.~\ref{hist_medobs} but for pixels
observed 240 times. \label{hist_lowobs} }
\end{figure}

It should be clear at this point that estimating iteratively the
noise properties yields no obvious advantage when compared to
the simpler WE and LPE described in this paper. Furthermore, as stressed 
in Sect.~1, the iterative scheme described above
needs a sound statistical foundation in the presence of long range
dependence. Moreover, it is important to
realize that such a scheme is severely more costly than LPE and WE. 
In fact, in order to implement the latter, only a
one step OLS map making is needed, plus
one periodogram estimation over the time line and the final GLS
one shot map making (the actual
computational effort to derive WE and LPE from
the periodogram is extremely small).
On the other hand, the iterative method requires, other than the
OLS map making, a periodogram estimation and a GLS 
map making at every noise iteration.

\subsection{Application to BOOMERanG}\label{app_to_boom}

All simulations shown above were performed for a subset of
the {\sc Planck} timestream. Furthermore, the noise model assumed
as input is the customary ``$1/f$'' noise (see 
Eq.~(\ref{planck_noise_model})).
In this Section we relax both assumptions and apply our
techniques to BOOMERanG.

BOOMERanG is a balloon borne experiment which scans at constant elevation
by allowing the gondola to slew at 1 or 2 degrees per second
(d.p.s.), while the detector sampling rate is 60~Hz (de Bernardis et 
al.\ 1999).
To make our simulations realistic we take a
chunk ($2^{22}$ samples  out of total of about $3 \cdot 10^7$) 
of the 1 d.p.s.\ part of the scan performed by the B150A channel
during the flight of 1998. 
Contrariously to {\sc Planck}, this scan is aimed
at a small patch. The central region
comprises a few tens of square degrees and its coverage
(i.e.\ number of hits per pixel) is far more uniform.
The noise properties of the BOOMERanG detectors are not well
described by the model given in Eq.~(\ref{planck_noise_model})
because the timestream is altered to deconvolve bolometric 
filters (both low pass and high pass). To take these
effects into account we propose to use:
\begin{equation}\label{noise_sim_boom}
P(f)=A\sin^2\left (\frac{\pi f}{2 f_0} \right ) 
\left [ 1+ \left (\frac{f_k}{f} \right)^{\alpha } + A^{\prime} 
e^{f/f_1}\right ]
\end{equation}
with fiducial values $A \simeq 2\cdot10^4\,(\muKrs)^2$,
$f_0 \simeq 10^{-3}$~Hz, $f_1 \simeq 7$~Hz,
$f_k = 0.15$~Hz, $A'\sim 10^8$ and $\alpha = 1.5$.
We choose $f_0 \ll f_1$, since
the exponential term arises after deconvonvolving the detector
high frequency response. To avoid excessive contamination from
this high frequency term we sharply cut the spectrum at 20~Hz
(as in BOOMERanG: see Hivon et al.\ 2001).
\begin{figure}
\resizebox{\hsize}{!}{\includegraphics{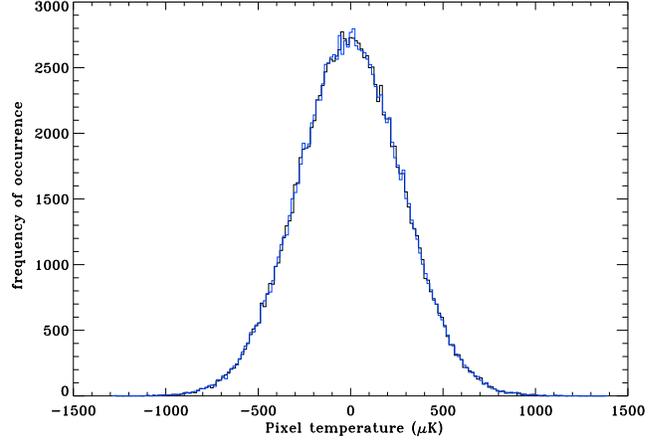}}
\caption{Same as Figs.~\ref{hist_medobs} and \ref{hist_lowobs},
but for a BOOMERanG-like simulation. The histogram are ``observed''
distributions of GLS mapmaking 
temperatures for all  pixel hit $23$ times in the maps. 
Noise has been estimated
using WE (black curve) and LPE (dark
blue).\label{hist_obs23_boom_long}}
\end{figure}
\begin{figure}
\resizebox{\hsize}{!}{\includegraphics{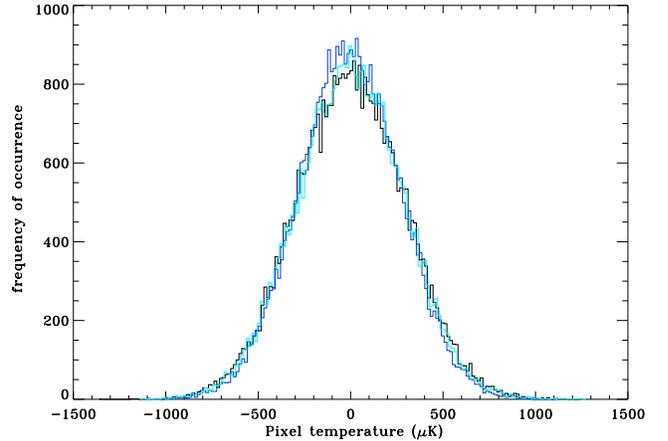}}
\caption{Same as Fig.~\ref{hist_obs23_boom_long} but for a pixel
hit 100 times. We compare here WE (dark blue curve) with the
iterative method (black).
The light blue curve is computed using the correct
noise power spectrum (Eq.~(\ref{noise_sim_boom})) and, hence,
provides the exact GLS estimate.\label{hist_obs23_boom}}
\end{figure}

We thus use Eq.~(\ref{srtsim}) to generate Gaussian noise assuming
the power spectrum given in Eq.~(\ref{noise_sim_boom}).
Subsequently, WE and LPE are applied to the data. 
WE is fully parametric and requires that a model
for the spectral density is specified. 
Rather than supply the model given by Eq.~(\ref{noise_sim_boom}), used to
generate the data and hence ``exact'', we base the estimations
on Eq.~(\ref{planck_noise_model}), i.e.\ on standard ``$1/f''$ noise. 
This results in a deliberate
model misspecification, as we are providing
WE with an incorrect model.
This choice has the obvious purpose to test the robustness
of the estimate. Strictly speaking, for $\alpha \le 2$ 
Eq.~(\ref{noise_sim_boom}) does no longer display long memory behavior,
and in this sense we are misleading LPE as well. However, long memory
is still a good approximation if $f_0 \ll f_k$.

Our results are shown in Fig.~\ref{hist_obs23_boom_long} where we give, 
for the most frequently observed pixel in our scan, temperature histograms
analogous to the ones given in Figs.~\ref{hist_medobs} and~\ref{hist_lowobs}.
The similarity between the two curves is striking.
Very much the same conclusions as in the previous Section can be
drawn: the difference between WE and LPE, and between WE and the iterative
method, is by all means negligible as shown in 
Fig.~\ref{hist_obs23_boom}. In the same Figure we plot the exact
(unfeasible) GLS estimate. Again, no significant difference emerges.

\section{Summary and conclusions}

This paper has addressed estimation methods for the properties of CMB
receiver noise which is well approximated by a long memory 
process. The most
recent literature proves that optimal statistical inference
is deeply affected by such characteristics. Under these circumstances
results by Dahlhaus (1995) entail that map making procedures
can be performed with a single iteration, provided that the slope at the
origin of the log spectral density (i.e. $\alpha $) is known; such
procedures are asymptotically equivalent to maximum likelihood, GLS-type
estimates. Our prime effort, hence, has been to address the efficient
estimation of the memory parameter $\alpha$. We have first presented the
fully parametric WE which allows for the full
correlation structure of the noise process to be recovered.
WE, however, requires that the functional form of the
noise spectral density is known. To relax this very restrictive assumption
we also focus on the semiparametric
case, where it is only assumed that a spectral singularity be present at the
origin. Both the
parametric and the semiparametric procedures are known
to enjoy some nice robustness properties in
non-standard situations, including non-Gaussian and non-stationary
circumstances. 

We assess the performance of these statistical procedures by implementing
them on simulated {\sc Planck} 
and BOOMERanG datasets. 
We note first that for practical implementation
we need to filter out the signal component by running a preliminary OLS
regression.
We have shown by direct Monte Carlo comparison that
this induces no significant degradation of the expected results.  
Our next conclusion is
that the performance of semiparametric LPE is very close to correctly
specified, fully parametric WE. This is comforting, as it allows for
robust statistical inference under minimal assumptions. In particular, in
terms of map making the improvement obtained by a full specification of the
noise structure is by all means negligible. Furthermore, our simple map
making procedure based on Dahlhaus (1995) ideas provides results which are
virtually indistinguishable from those produced by the (computationally much
more costly) iterative scheme of Prunet et al.\ (2001). These results are
robust to implementation in the stationary and nonstationary regions.

We believe this paper leaves several avenues for further research.
For example, 
distribution (goodness of fit) tests are profoundly affected by long memory
behaviour (Dehling \& Taqqu, 1989; Giraitis \& Surgailis, 1999).
Efficient estimation of $\alpha $ can therefore be exploited to investigate
in some detail non-Gaussian features. Also,
a correct evaluation of the dependence structure of receiver
noise may in principle improve the determination of confidence intervals for
the angular power spectrum and its governing parameters, either by
resampling methods or by asymptotic approximations.
These issues are currently being investigated.

\begin{acknowledgements} 
We thank P.\ de Bernardis and the BOOMERanG
collaboration for having provided us with the BOOMERanG scan.
We acknowledge  use of the HEAL\-Pix 
package\footnote{current web\-site: http://www.eso.org/science/healpix} 
(Gor\-ski et al.\ 1999)
and of the FFTW library (Frigo \& Johnson, 1998). 
\end{acknowledgements}


\begin{thebibliography}{}


\bibitem[Balbi et al., 2000]{balbi} Balbi, A.\ et al., 2000, 545, \apj, L1

\bibitem[Bond et al.\ 1998]{Bond_spectrum}
Bond, J.R., Jaffe, A.H., \& Knox, L., 1998, \PR\
D57, 2117, 1998

\bibitem[Borrill 1999]{borrill} Borrill, J, 1999, Proceedings
of the Conference ``{\it 3~K Cosmology},'' AIP Conf.\  Proc.\ 476, 277

\bibitem[Dahlhaus, 1989]{dah} Dahlhaus, R., 1989, \Anst, 17, 1749

\bibitem[Dahlhaus, 1995]{dah2} Dahlhaus, R., 1995, \Anst, 23, 1029

\bibitem[de Bernardis et al., 1999]{debbe1} de Bernardis, P.\ et al., 1999,
New Astr.\ Rev., 43, 289

\bibitem[de Bernardis et al., 2001]{debbe2} de Bernardis, P.\ et al., 2001,
\apj\ in press [astro-ph/0105296]

\bibitem[Dehling \& Taqqu, 1989]{dehling_and_taqqu} 
Dehling, H.\ \& Taqqu, M.S., 1989, \Anst, 17, 1767 

\bibitem[Dor\'e et al., 2001]{dore} Dor\'e, O., 
Teyssier, R., Bouchet, F.R., Vibert, D. \&
Prunet, S., 2001, \AsAs, 374, 358  

\bibitem[Ferreira \& Jaffe, 1989], Ferreira, P.G.F. \& Jaffe, A.H.,
2000, \MNRAS, 312, 89
 
\bibitem[Fox]{fox_taqqu} Fox, R.\ \& Taqqu, M., 1986, \Anst, 14, 517

\bibitem[Frigo \& Johnson 1998]{fftw} Frigo, M. \& Johnson, S.G., 1998 
ICASSP Conference, 3, 1381; also see \verb+http://www.fftw.org/+

\bibitem[Geweke and]{gew_ph} Geweke, J.\ \& Porter-Hudak, S., 1983
J.\ Time Ser.\ Anal.\ 4, 221

\bibitem[Giraitis and Surgalis]{gir_surg} Giraitis, L. \& Surgailis, D.,
1990, Probab.\ Theory Relat.\ Fields, 86, 87

\bibitem[Giraitis and Surgalis]{girs_surg3} Giraitis, L. \& Surgailis, D., 
1999, J.\ Statist.\ Plann.\ Inference, 80 81. 

\bibitem[Giraitis and Taqqu]{gir_surg2} Giraitis, L. \& Taqqu, M.,
1999, \Anst, 27, 178

\bibitem[G\'orski et al.\ 1998]{gorski_healpix}
G\'orski, K.M., Hivon, E. \& Wandelt, B.D., 1999 
in ``Evolution of large scale structure: from recombination to 
Garching'', ed.\ by A.J.\ Banday, R.K.\ Sheth, L.N.\ da Costa,
proc.\ of the MPA-ESO Cosmology conference, Garching, Germany, 
2-7 August 1998, 37, PrintPartners IPSKAMP NL [astro-ph/9812350]

\bibitem[Hivon et al. 2001]{hivon} Hivon, E.\ et al., 2001, \apj\ in press
[astro-ph/0105302]

\bibitem[Mouline \& Souliers]{moul_soul} Moulines, E. \& 
Soulier, P., 1999, \Anst, 27, 1415

\bibitem[Natoli et al. 2001]{natoli} Natoli P., 
de Gasperis, G., Gheller, C.\ \& Vittorio, N., 2001, \AsAs, 372, 346

\bibitem[Oh, Spergel \& Hinshaw 1999]{osh} Oh, S.P., 
Spergel, D., Hinshaw, G., 1999, \apj, 510, 551  

\bibitem[Press \etal\ 1992]{nr} Press, W. H., Flannery, B. P.,
Teukolsky, S. A.  \&
Vetterling, W. T., 1992, Numerical Recipes in FORTRAN, The Art of
Scientific Computing, $2^{nd}$ Edition, Cambridge University
Press, Cambridge

\bibitem[Prunet et al., 2001]{prunet} Prunet et al., 2001, to appear
in proc.\ of the MPA/ESO conference ``Mining the Sky'' 
[astro-ph/0101073]  

\bibitem[Velasco]{velasco} Velasco, C., 1999, J.\ Econom., 91, 325

\bibitem[Velasco \& Robinson]{velasco_rob}  Velasco, C.\ \& Robinson,
P.M., 2000, J.\ Amer.\ Statist.\ Assoc., 452, 1229

\bibitem[Robinson]{robinson_a} Robinson, P.M., 1995a,  \Anst, 23, 1048

\bibitem[Robinsonb]{robinson_b}  Robinson, P.M., 1995b,  \Anst, 23, 1630

\bibitem[Robinson \& Marinucci]{rob_mar} Robinson, P.M.\ \& Marinucci, D.,
2001,  \Anst\ in press

\bibitem[Tegmark 1997]{Tegmark_mapmaking} Tegmark, M., 1997, \PR, D55, 5895

\bibitem[whittle]{whittle} Whittle, P., 1953,  Ark.\ Mat.\ 2, 423 

\bibitem[Wright 1996]{wright96} Wright, E.L., 1996, paper presented at the 
IAS CMB Data Analysis Workshop [astro-ph/9612006]

\end{thebibliography}
\end{document}